\documentclass[twocolumn,aps,prd,showpacs,10pt,superscriptaddress]{revtex4-1}

\usepackage{amssymb}
\usepackage{amsmath}
\usepackage{bm}
\usepackage{multirow}
\usepackage{graphicx}
\usepackage[usenames,dvipsnames,svgnames]{xcolor}
\usepackage{hyperref}
\usepackage[normalem]{ulem}
\hypersetup{
pdfnewwindow=true,      
colorlinks=true,        
linkcolor=Blue,         
citecolor=Blue,         
filecolor=Blue,         
urlcolor=Blue           
}

\begin{document}


\title{Interpretation of apparent thermal conductivity in finite systems from equilibrium molecular dynamics simulations}

\author{Haikuan Dong}
\affiliation{Beijing Advanced Innovation Center for Materials Genome Engineering, University of Science and Technology Beijing, Beijing, 100083, China}
\affiliation{MSP group, QTF Centre of Excellence, Department of Applied Physics, Aalto University, FI-00076 Aalto, Finland}
\affiliation{College of Physical Science and Technology, Bohai University, Jinzhou, 121013, China}
\author{Shiyun Xiong}
\affiliation{Functional Nano and Soft Materials Laboratory (FUNSOM) and Collaborative Innovation Center of Suzhou Nano Science and Technology, Soochow University, 215123 Suzhou, China}
\author{Zheyong Fan}
\email{brucenju@gmail.com}
\affiliation{MSP group, QTF Centre of Excellence, Department of Applied Physics, Aalto University, FI-00076 Aalto, Finland}
\affiliation{College of Physical Science and Technology, Bohai University, Jinzhou, 121013, China}
\author{Ping Qian}
\email{qianping@ustb.edu.cn}
\affiliation{Beijing Advanced Innovation Center for Materials Genome Engineering, University of Science and Technology Beijing, Beijing, 100083, China}
\affiliation{Department of Physics, University of Science and Technology Beijing, Beijing 100083, China}
\author{Yanjing Su}
\email{yjsu@ustb.edu.cn}
\affiliation{Beijing Advanced Innovation Center for Materials Genome Engineering, University of Science and Technology Beijing, Beijing, 100083, China}
\affiliation{Corrosion and Protection Center, University of Science and Technology Beijing, Beijing, 100083, China}
\author{Tapio Ala-Nissila}
\affiliation{MSP group, QTF Centre of Excellence, Department of Applied Physics, Aalto University, FI-00076 Aalto, Finland}
\affiliation{Interdisciplinary Centre for Mathematical Modelling, Department of Mathematical Sciences, Loughborough University, Loughborough, Leicestershire LE11 3TU, UK}

\date{\today}

\begin{abstract}
We propose a way to properly interpret the apparent thermal conductivity obtained for finite systems using equilibrium molecular dynamics simulations (EMD) with fixed or open boundary conditions in the transport direction. In such systems the heat current autocorrelation function develops negative values after a correlation time which is proportional to the length of the simulation cell in the transport direction. Accordingly, the running thermal conductivity develops a maximum value at the same correlation time and eventually decays to zero. By comparing EMD with nonequilibrium molecular dynamics (NEMD) simulations, we conclude that the maximum thermal conductivity from EMD in a system with domain length $2L$ is equal to the thermal conductivity from NEMD in a system with domain length $L$. This facilitates the use of nonperiodic-boundary EMD for thermal transport in finite samples in
close correspondence to NEMD.
\end{abstract}

\maketitle

\section{Introduction}

Nonequilibrium molecular dynamics (NEMD) \cite{hoover1975tcap,ciccotti1980jsp,Tenenbaum1982pra,Mountain1983prb,ikeshoji1994mp,muller-plathe1997jcp,jund1999prb,shiomi2014arht}  has been the standard method for computing the \textit{apparent} or {\it effective} thermal conductivity $\kappa(L)$ of a finite system of length $L$ in the transport direction. For good thermal conductors, $\kappa(L)$ depends on $L$ at the nanometer scale or even the micrometer scale, depending on the average phonon mean free path in the bulk system. On the other hand, equilibrium molecular dynamics (EMD), where the thermal conductivity is calculated as a time integral of the heat current autocorrelation function (HCACF) according to a Green-Kubo relation \cite{green1954jcp,kubo1957jpsj}, is the standard method for computing the bulk thermal conductivity in the thermodynamic limit. In the EMD method, periodic boundary conditions are applied to the transport direction, and the calculated thermal conductivity is regarded as that for an infinitely long system, although one has to be mindful for possible finite-size effects introduced by the use of a finite simulation domain. In this regard, the simulation domain size in EMD does not correspond to that in NEMD, and is therefore not related to a physical sample size in experiments. 
Therefore, it has been concluded that EMD cannot be used to calculate the apparent thermal conductivity of finite systems, although EMD simulations have been used for computing interfacial thermal conductance \cite{puech1986jltp,barrat2003mp,mcgaughey2006asme,chalopin2012prb,merabia2012prb,liang2014prb}.  For example, this view has been explicitly expressed by Matsubara \textit{et al.} \cite{Matsubara2020drm} in a study of the thermal transport properties of diamond nanoparticles. Because the thermal conductivity computed from EMD simulations is that for an infinitely large (periodic) system, previous works have been focused on evaluating the equivalence between the converged Green-Kubo integral and the NEMD results extrapolated to the limit of infinite system length \cite{schelling2002prb,sellan2010prb,he2012pccp,dong2018prb}.

In this work, we show that with appropriate modifications to the boundary conditions in the transport direction, the EMD method {\it can} actually be used to obtain the apparent thermal conductivity of finite systems. Specifically, instead of using periodic boundary conditions in the transport direction, we use fixed or open boundary conditions. In these cases, the \textit{running} thermal conductivity will first increase with increasing correlation time, but will eventually decay to zero, developing a maximum value at a particular correlation time. We show that this maximum value of the thermal conductivity for a domain length of $2L$ is the same as that obtained from an NEMD with a simulation domain length of $L$. This validates the use of nonperiodic-boundary EMD to study thermal transport in finite-size samples.

\section{Models and Methods}

\begin{figure*}[htb]
\begin{center}
\includegraphics[width=2\columnwidth]{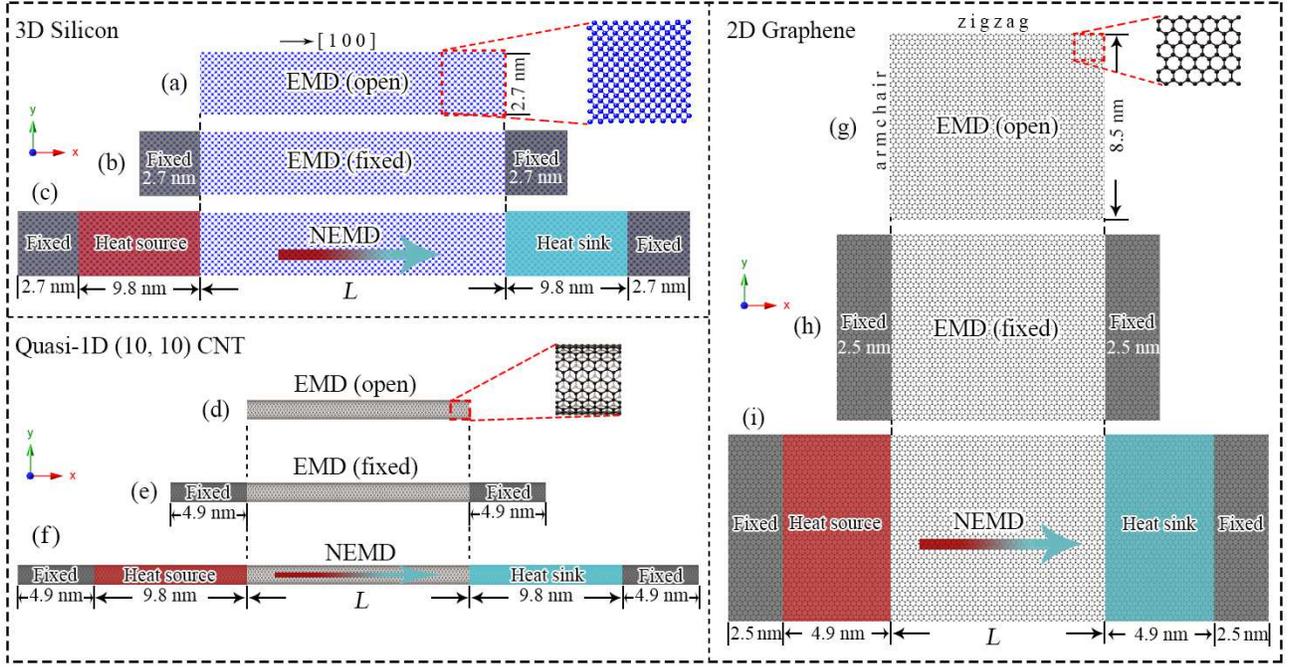}
\caption{Schematic illustration of the setups for the different MD simulations: (a)-(c) EMD with open boundary conditions in the transport direction, EMD with fixed boundary conditions in the transport direction ([100] direction), and NEMD with fixed boundary conditions in the transport direction, respectively, for 3D silicon. (d)-(f) similarly for quasi-1D $(10, 10)$ CNT, with the transport being along the tube. (g)-(i) similarly for 2D graphene, with the transport being along the zigzag direction.}
\label{model_si}
\end{center}
\end{figure*}

To demonstrate the general applicability of our results, we consider materials in different dimensions, including a 3D silicon crystal, 2D graphene, and a quasi-1D carbon nanotube (CNT) with chirality $(10,10)$. These materials are of great technological importance in the context of thermal management and thermoelectric energy conversion, and their thermal transport properties have attracted great attention in the past \cite{Rurali2010RMP,marconnet2013rmp,Gu2018RMP}. The setups for our EMD and NEMD simulations are schematically illustrated in Fig. \ref{model_si}. 

Figures \ref{model_si}(a), \ref{model_si}(d) and \ref{model_si}(g) show the EMD simulation setup with open boundary or free conditions in the transport direction. 
Figures \ref{model_si}(b), \ref{model_si}(e), and \ref{model_si}(h) show the EMD simulation setup with fixed boundary conditions in the transport direction, where some extra atoms at the two ends in the transport direction are fixed (frozen). In all these cases, the unfixed part has a length of $L$, which we call the simulation domain length. After thermal equilibration, the system is evolved in the $NVE$ ensemble and the equilibrium heat current 
\begin{equation}
    \bm{J}= \sum_i\sum_j \bm{r}_{ij} \frac{\partial U_j}{\partial \bm{r}_{ji}} \cdot \bm{v}_i
\end{equation}
is sampled. Here, $U_j$ is the site potential of atom $j$, $\bm{v}_i$ is the velocity of atom $i$, and $\bm{r}_{ij}=-\bm{r}_{ji} = \bm{r}_j - \bm{r}_i$ is the relative position from atom $i$ to atom $j$. For a derivation of the heat current formula and definition of the site potential, see Ref. \cite{fan2015prb}. From the sampled heat current one can then calculate the HCACF $\langle J_{x}(0) J_{x}(\tau)\rangle$ (taking $x$ as the transport direction) and the running, time dependent thermal conductivity through the following Green-Kubo relation  \cite{green1954jcp,kubo1957jpsj}:
\begin{equation}
\kappa(\tau) = \frac{1}{k_{\rm B} T^2 V}\int_0^{\tau} \langle J_{x}(0) J_{x}(t) \rangle dt.
\end{equation}
Here, $k_{\rm B}$ is Boltzmann's constant, $T$ is the system temperature, and $V$ is the system volume. For graphene and CNT, a conventional effective thickness of $0.335$ nm for the carbon layer is chosen to calculate the volume. 

\begin{figure}[htb]
\begin{center}
\includegraphics[width=1\columnwidth]{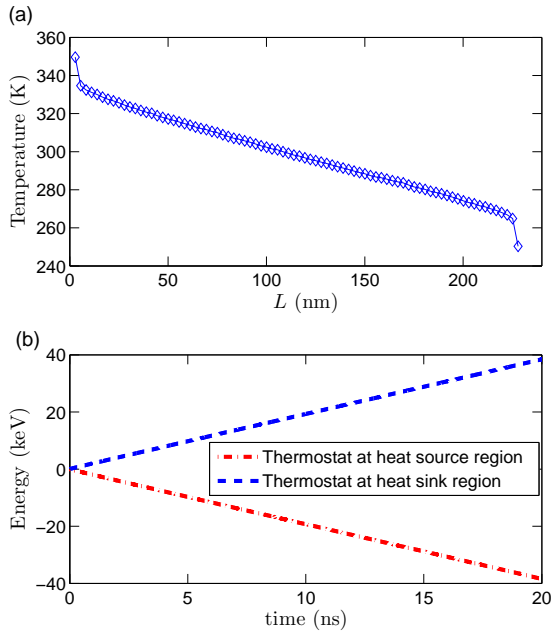}
\caption{ (a) Steady-state temperature profile and (b) energies of the thermostats coupled to the heat source and sink regions as a function of the time. The heat transfer rate $dE/dt$ is calculated as the average of the absolute slopes in the two curves. The energy in the thermostat coupled to the heat source (sink) region is decreasing (increasing) because it releases (absorbs) energy to maintain the higher (lower) temperature in the thermostatted region. }
\label{figure:nemd_diagram}
\end{center}
\end{figure}

Figures \ref{model_si}(c), \ref{model_si}(f), and \ref{model_si}(i) show the NEMD simulation setup with fixed boundary conditions in the transport direction, where apart from some extra fixed atoms at the two ends in the transport direction, there are also two thermostatted regions between the fixed atoms and the middle part that defines the sample length $L$. One thermostatted region is maintained at a higher temperature $T+\Delta T/2$ (corresponding to a heat source) and the other is at a lower temperature $T-\Delta T/2$ (corresponding to a heat sink), inducing a directional nonequilibrium heat current as indicated by the arrow within the sample region. Following Refs. \cite{li2019jcp,hu2020prb}, the Langevin thermostat with a coupling time of $0.1$ ps is used to generate the heat source and sink. The apparent thermal conductivity in the transport direction is calculated as \cite{li2019jcp}
\begin{equation}
\label{equation:nemd}
\kappa(L) = \frac{dE/dt}{A\Delta T / L},
\end{equation}
where $A$ is the cross-sectional area in the transverse directions and $dE/dt$ is the average energy exchange rate between the thermostats and the thermostatted regions. A temperature profile and the corresponding accumulative energies in the thermostats are shown in Fig. \ref{figure:nemd_diagram}.

All the EMD and NEMD simulations were performed using the GPUMD package \cite{fan2017cpc}. For 3D silicon crystal, we used the minimal Tersoff potential \cite{fan2019jpcm}. For both 2D graphene and the quasi-1D $(10,10)$ CNT, we used the Tersoff potential \cite{tersoff1989prb} parameterized by Lindsay and Broido \cite{lindsay2010prb}. A time step of $1$ fs was used in simulations at $300$ K, and for a series of simulations for the $(10,10)$ CNT from 300 K to 1300 K, the time step was decreased from $1$ fs to $0.1$ fs. For both EMD and NEMD simulations, periodic boundary conditions were applied in the two transverse directions (of area of about $2.7\times 2.7$ nm$^2$) for 3D silicon crystal and the transverse direction (of width of about $8.5$ nm) in the basal plane for 2D graphene. The lengths of the fixed regions and thermostatted regions for all the models can be found in Fig. \ref{model_si}. The domain lengths considered here (indicated as $L$ in each panel of Fig. \ref{model_si}) for the NEMD simulations are as follows: $L$ = $13.6$, $27.2$, $54.3$, $108.6$, and $217.2$ nm for 3D silicon crystal, $L$ = $12.3$, $24.6$, $49.2$, $98.4$, and $196.8$ nm for 2D graphene, and $L$ = $24.6$, $49.2$, $98.4$, $196.8$, and $393.5$ nm for quasi-1D $(10,10)$ CNT. For the EMD simulations, the domain lengths are twice as large. The reason for this choice will be mentioned later. In the EMD simulations, we performed $10$ independent runs for each domain length, each with a production time of $10$ ns. In the NEMD simulations, we performed $3$ independent runs for each domain length, each with a production time of $10$ ns. Error bounds of the presented data were calculated as standard errors, i.e., standard deviations divided by the number of independent runs.

\section{Results and discussion}

\begin{figure}[htb]
\begin{center}
\includegraphics[width=\columnwidth]{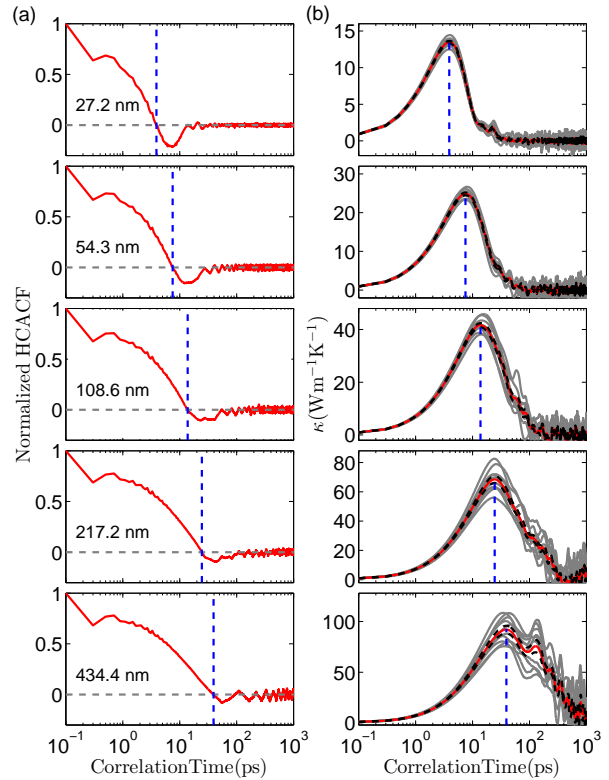}
\caption{(a) Normalized HCACF and (b) running thermal conductivity $\kappa(\tau)$ as a function of the correlation time for different sample lengths $L$ from EMD simulations with open boundary conditions in the transport direction. For the running thermal conductivity, the red thick line represents the mean values from ten independent runs (the gray thin lines) and the black dashed lines indicate the standard error. For the HCACF, only the average values are shown for clarity. The blue dashed vertical lines  correspond to the time $\tau_{\rm max}$ at which $\kappa(\tau)$ reaches a maximum, and equivalently after which the HCACF develops negative values. The systems here are 3D silicon crystals at $300$ K and zero pressure, with the length $L$ for each sample written in the corresponding panel.}
\label{figure:emd_open}
\end{center}
\end{figure}

\begin{figure}[htb]
\begin{center}
\includegraphics[width=\columnwidth]{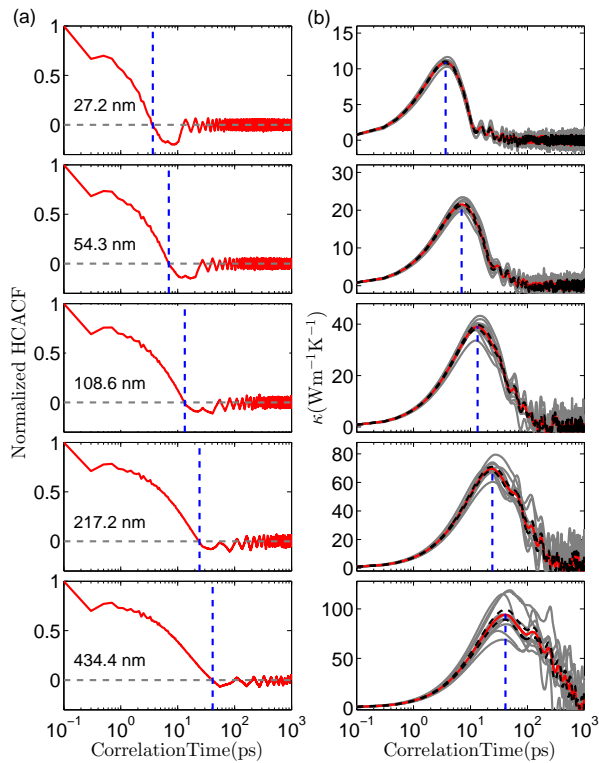}
\caption{Similar to Fig. \ref{figure:emd_open}, but for EMD simulations with fixed boundary conditions, instead of open boundary conditions, in the transport direction.}
\label{figure:emd_fixed}
\end{center}
\end{figure}

We first consider a 3D silicon crystal. The normalized HCACF $\langle J_{x}(0) J_{x}(\tau)\rangle$ and running thermal conductivity $\kappa(\tau)$ as a function of the correlation time $\tau$ from EMD simulations with open boundary conditions in the transport direction are presented in Fig. \ref{figure:emd_open}. Figure \ref{figure:emd_fixed} shows the results from EMD simulations with fixed boundary conditions in the transport direction. For each sample length, the normalized HCACF first decreases with increasing $\tau$, then changes from positive to negative at a particular correlation time $\tau=\tau_{\rm max}$, and finally decays to zero from the negative side. Accordingly, the running thermal conductivity $\kappa(\tau)$ first increases with increasing $\tau$, then develops a maximum value at $\tau_{\rm max}$, and eventually decays to zero from the positive side, with fluctuations in the long-time limit due to increasing noise-to-signal ratio. Such peaks in the running thermal conductivity have been observed in other contexts, such as thermal transport in nanoporous silicon \cite{Oliveira2020prb} and nonlocal thermal transport within the linear-response formalism \cite{Fernando2020arxiv}.

\begin{figure}[htb]
\begin{center}
\includegraphics[width=\columnwidth]{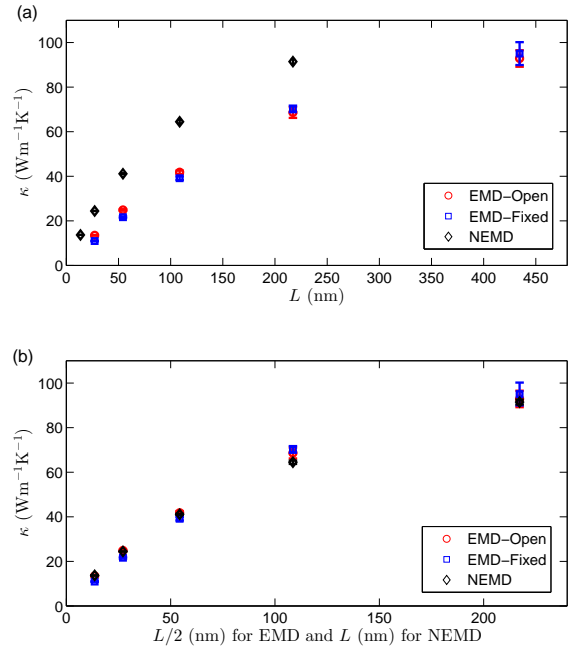}
\caption{(a) The maximum thermal conductivity $\kappa_{\rm max}$ from EMD simulations with open and fixed boundary conditions in the transport direction and the apparent thermal conductivity from NEMD simulations as a function of the domain length $L$ for 3D crystal silicon at 300 K and zero pressure. (b) Similar to (a) but using $L/2$ as the horizontal axis for the EMD data.}
\label{figure:emd_vs_nemd_si}
\end{center}
\end{figure}

Both the HCACF and the running thermal conductivity here are very different from those from conventional EMD simulations with periodic boundary conditions applied in the transport direction, where $\kappa(\tau)$ converges to a finite value (regarded as the thermal conductivity for an infinitely long system if finite-size effects are eliminated) instead of zero. The fact that $\kappa(\tau)$ converges to zero in the EMD simulations with open or fixed boundary conditions in the transport direction has led Matsubara \textit{et al.} \cite{Matsubara2020drm} to conclude that the Green-Kubo relation cannot be used to compute the thermal conductivity for finite systems. However, as $\kappa(\tau)$ has a well defined maximum value at $\tau_{\rm max}$, it is reasonable to conjecture that this maximum value is related to the apparent thermal conductivity for a finite system as computed from an NEMD simulation.

To explore this conjecture, we first compute the maximum $\kappa$ values for the five samples in the EMD simulations and plot them in Fig. \ref{figure:emd_vs_nemd_si}(a) as a function of the domain length $L$. These values are compared against the $\kappa(L)$ values computed from the NEMD simulations according to Eq. (\ref{equation:nemd}). From Fig. \ref{figure:emd_vs_nemd_si}(a), we see that $\kappa$ increases with increasing $L$ in both EMD and NEMD simulations, but their values do not match for each $L$. However, remarkably enough if we use $L/2$ as the horizontal axis for data from EMD simulations, the EMD and NEMD data become mutually consistent as can be seen from Fig. \ref{figure:emd_vs_nemd_si}(b). This quantitative comparison suggests a clear relation: The maximum thermal conductivity from EMD simulations with open or fixed boundary conditions in the transport direction in a system with domain length $2L$ equals to the apparent thermal conductivity from NEMD simulations in a system with domain length of $L$:
\begin{equation}
\label{equation:nend_emd}
\kappa^{\rm EMD}_{\rm max}(2L) = \kappa^{\rm NEMD}(L).
\end{equation}

\begin{figure}[htb]
\begin{center}
\includegraphics[width=\columnwidth]{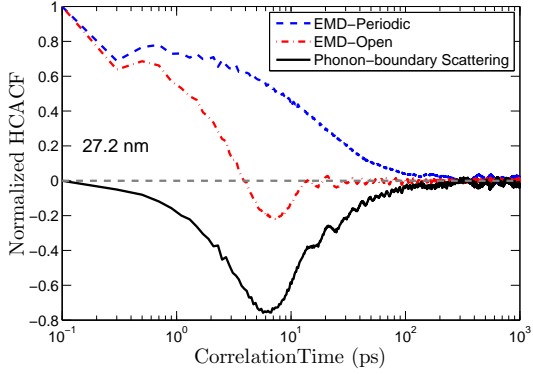}
\caption{Normalized HCACF in silicon crystal (300 K and zero pressure) in different conditions. The blue dashed line represents the HCACF obtained from EMD simulations with periodic boundary conditions in the transport direction. The red dot-dashed line represents the HCACF obtained from EMD simulations with open boundary conditions in the transport direction. The black solid line represents the difference between the above two, which is induced by phonon-boundary scattering.}
\label{figure:hcacf}
\end{center}
\end{figure}

\begin{figure}[htb]
\begin{center}
\includegraphics[width=\columnwidth]{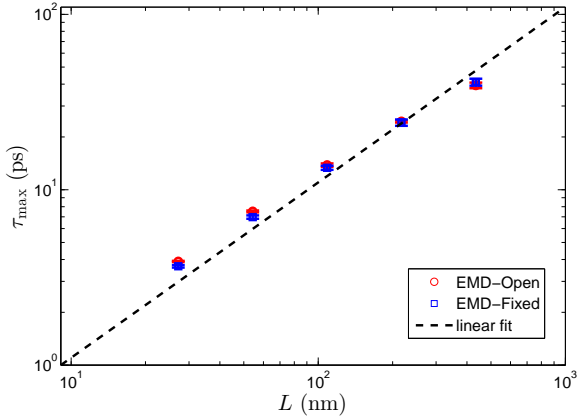}
\caption{The correlation time $\tau_{\rm max}$ at which the running thermal conductivity $\kappa(\tau)$ attains the maximum value $\kappa_{\rm max}$ as shown in Fig. \ref{figure:emd_open} and Fig. \ref{figure:emd_fixed} against the EMD simulation domain length $L$.}
\label{figure:time_vs_L}
\end{center}
\end{figure}

\begin{figure}[htb]
\begin{center}
\includegraphics[width=\columnwidth]{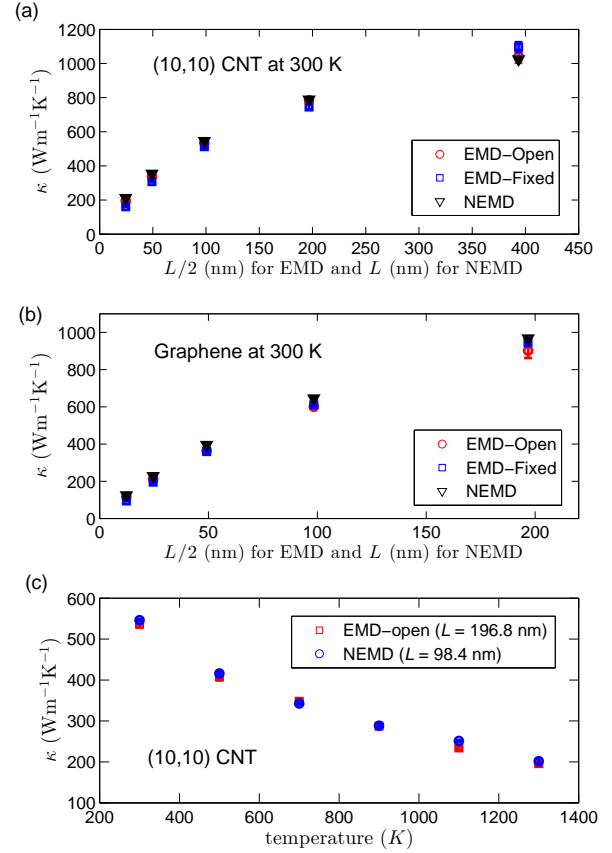}
\caption{Validation of Eq. (\ref{equation:nend_emd}) for (a) $(10,10)$ CNT with different lengths, (b) graphene sheet with different lengths, and (c) $(10,10)$ CNT with fixed lengths but at different temperatures $T$.}
\label{figure:cnt_gra}
\end{center}
\end{figure}

The physical explanation behind this unexpected relation can be found as follows. The reason why $\kappa(\tau) \to 0$ as $\tau \to \infty$ and the existence of a maximum value of $\kappa(\tau)$ at a particular correlation time $\tau_{\rm max}$ originate from boundary scattering of the phonons in the system. In the context of the Green-Kubo relation, or equivalently the fluctuation-dissipation theorem, boundary scattering will induce a negative HCACF due to forward (towards the boundary) and backward (reflected from the boundary) heat currents generated by spontaneous fluctuations at equilibrium. This negative HCACF is shown in Fig. \ref{figure:hcacf}. It is computed as the difference between the HCACF obtained from EMD simulations with periodic boundary conditions (with no boundary scattering) and the HCACF obtained from EMD simulations with open boundary conditions in the transport direction. The magnitude of the negative HCACF increases from zero to a maximum value at a time when, on average, the forward and backward heat currents meet each other. This time should be close to $\tau_{\rm max}$ after which the total HCACF (due to both phonon-phonon scattering and phonon boundary scattering) becomes negative.

In the quasi-ballistic regime, heat currents (or heat waves) propagate at a speed of the order of the phonon group velocity $v_{\rm g}$, and the average time for a forward heat wave to meet the backward heat wave is $L/v_{\rm g}$, and we therefore have $\tau_{\rm max} \approx L/v_{\rm g}$. This relation is confirmed in Fig. \ref{figure:time_vs_L}. Based on the linear fit in Fig. \ref{figure:time_vs_L}, we can estimate $v_{\rm g}$ to be about $10$ km/s, which is a reasonable value for silicon crystal.

The arguments above explain the development of a maximum thermal conductivity $\kappa_{\rm max}$ at a correlation time $\tau_{\rm max}$ due to phonon-boundary scattering. These arguments can also be used to understand the quantitative relation between $\kappa_{\rm max}$ from EMD and the apparent thermal conductivity from NEMD, as expressed in Eq. (\ref{equation:nend_emd}). It is well known that the length dependence of the apparent thermal conductivity from NEMD is also caused by phonon-boundary scattering. However, there is a difference in the mean free path of phonon-boundary scattering events for our EMD and NEMD setups as shown in Fig. \ref{model_si}. In the NEMD simulations, phonons are released from the heat source region and absorbed in the heat sink region, and the mean free path due to phonon-boundary scattering is $L$. In the EMD simulations, on the other hand, phonons generated by spontaneous fluctuations can only propagate a distance of $L/2$ on average before experiencing boundary scattering, and the mean free path due to phonon-boundary scattering is thus $L/2$. This naturally explains the relation in Eq. (\ref{equation:nend_emd}).

The previous results were obtained for a silicon crystal. To show that our results apply in general, we also consider other materials of different dimensions, including 2D graphene and quasi-1D $(10,10)$-CNT. Figures \ref{figure:cnt_gra}(a) and \ref{figure:cnt_gra}(b) show that Eq. (\ref{equation:nend_emd}) also holds for these systems. We also considered the $(10,10)$-CNT with a fixed length $L$ at a wide range of temperatures from $300$ K to $1300$ K. Figure \ref{figure:cnt_gra}(c) shows that Eq. (\ref{equation:nend_emd}) is valid for all the temperatures considered. These extensive MD simulations suggest that Eq. (\ref{equation:nend_emd}) is valid in general as our boundary scattering argument suggests.

\section{Summary and Conclusions}

In summary, we have explored the physics underlying EMD simulations with nonperiodic (open or fixed) boundary conditions in the transport direction. In this case, the heat current autocorrelation function develops negative values after a particular correlation time $\tau_{\rm max}$, at which the running thermal conductivity from the Green-Kubo integral attains a maximum value $\kappa_{\rm max}$. Based on extensive EMD and NEMD simulations of materials with different spatial dimensions, lengths, and temperatures, we have found the unexpected result that $\kappa_{\rm max}$ from nonperiodic-boundary EMD simulations with a domain length of $2L$ equals the apparent thermal conductivity $\kappa(L)$ from NEMD simulations with a domain length of $L$. The physical origin of this result comes from the fact that the mean-free path induced by phonon-boundary scattering in the nonperiodic-boundary EMD simulations is only half of the simulation domain length, while it corresponds to the full domain length in the NEMD simulations.

\begin{acknowledgments}
 This work was supported by the National Key Research and Development Program of China under Grant Nos. 2016YFB0700500 and 2018YFB0704300, the National Natural Science Foundation of China under Grant No. 11974059 and 11804242, and the Academy of Finland through its QTF Centre of Excellence Programme under project No. 312298. We acknowledge the computational resources provided by Aalto Science-IT project and Finland's IT Center for Science (CSC).
\end{acknowledgments}

\end{document}